\documentclass[twocolumn,showpacs,preprintnumbers,aps,prl,amsmath,amssymb,superscriptaddress]{revtex4}

\usepackage{color}
\usepackage{graphicx}
\usepackage{dcolumn}
\usepackage{bm}
\usepackage{amssymb}
\usepackage[german, english]{babel}
\usepackage{dcolumn} %align table columns on decimal point
\begin{document}
\title{Collisional Stability of $^{40}$K Immersed in a Strongly Interacting Fermi Gas of $^6$Li}
\author{F. M.\ Spiegelhalder}
\author{A.\ Trenkwalder}
\author{D.\ Naik}
\author{G.\ Hendl}
\author{F.\ Schreck}
 \affiliation{Institut f\"ur Quantenoptik und Quanteninformation,
\"Osterreichische Akademie der Wissenschaften, 6020 Innsbruck,
Austria}
\author{R.~Grimm}
\affiliation{Institut f\"ur Quantenoptik und Quanteninformation,
\"Osterreichische Akademie der Wissenschaften, 6020 Innsbruck,
Austria}
 \affiliation{Institut f\"ur Experimentalphysik und
Zentrum f\"ur Quantenphysik, Universit\"at Innsbruck,
6020 Innsbruck, Austria}

\date{\today}

% 67.85.-d Ultracold gases, trapped gases  (see also 03.75.-b Matter waves in quantum mechanics)
% 34.50.-s Scattering of atoms and molecules
% 05.30.Fk Fermion systems and electron gas  (see also 71.10.-w Theories and models of many-electron systems; see also 67.10.Db Fermion degeneracy in quantum fluids)
% 67.85.Lm degenerate Fermi gases

\begin{abstract}
We investigate the collisional stability of a sample of $^{40}$K atoms immersed in a tunable spin mixture of $^6$Li atoms. In this three-component Fermi-Fermi mixture, we find very low loss rates in a wide range of interactions as long as molecule formation of $^6$Li is avoided. The stable fermionic mixture with two resonantly interacting spin states of one species together with another species is a promising system for a broad variety of phenomena in few- and many-body quantum physics.
\end{abstract}

\maketitle

The groundbreaking achievements in experiments with ultracold Fermi gases have opened up unprecedented possibilities to study new regimes of strongly interacting quantum matter \cite{Inguscio2006ufg, Giorgini2008tou, Bloch2008mbp}.
Recent experiments have opened up two important new research frontiers with fermionic atoms that go beyond the two-component spin mixtures so far exploited in the field. Mixtures involving three different spin states \cite{Ottenstein2008cso, Huckans2009tbr} and mixtures of different fermionic species \cite{Taglieber2008qdt, Wille2008eau, Tiecke2009sfb} have produced first exciting results like the demonstration of fermionic Efimov states \cite{Wenz2009aut, Williams2009efa} and the creation of Fermi-Fermi molecules \cite{Voigt2009uhf, moredetails}.

A key ingredient in the success story of strongly interacting Fermi gases is their collisional stability. In contrast to other systems with large scattering lengths, which usually exhibit very fast collisional decay, a two-component Fermi gas can be extraordinarily stable. This results from Pauli suppression effects in atomic three-body decay \cite{Esry2001tlf} and in atom-dimer and dimer-dimer collisions \cite{Petrov2004wbm}. Obviously, such a suppression effect is absent when three distinguishable particles interact. This raises general questions concerning collisionally stable regimes and thus the possibilities for experiments on multicomponent Fermi gases with strong interactions. The recent experiments \cite{Ottenstein2008cso, Huckans2009tbr} on three-component spin mixtures of $^6$Li have indeed shown very rapid collisional decay in regimes involving large scattering lengths.

In this Letter, we experimentally explore a three-component Fermi gas with strong interactions between two of its components. A single spin state of $^{40}$K is immersed in a deeply degenerate two-component spin mixture of $^6$Li atoms. While the intraspecies interaction between the two $^6$Li spin states can be tuned to arbitrarily large strength, the interspecies interaction remains weak. We find that a wide stable region exists where elastic collisions dominate over inelastic decay. This opens up a variety of new exciting experimental possibilities. Here we demonstrate two immediate applications, the sympathetic cooling of another species by a strongly interacting Fermi gas and the possibility of precise thermometry in the strongly interacting regime.

The Fermi-Fermi mixture is prepared in an optical dipole trap \cite{moredetails}. Initially about $10^7$ Li atoms and a few $10^4$ K atoms are loaded from a two-species magneto-optical trap (MOT) into the focus of an intense infrared laser beam (initial power 70\,W, wavelength 1070\,nm, waist $25\,\mu$m). The spin states are prepared such that Li is in an equal mixture of its two lowest energy states, labeled Li$|1\rangle$ and Li$|2\rangle$, and K is in its lowest state, labeled K$|1\rangle$. After a forced evaporative cooling process, which takes typically 10\,s, the atoms are trapped in a 75\,mW beam (waist $\sim$45$\mu$m). In the final evaporation phase, the axial trapping is essentially provided by the magnetic-field curvature of $28\,$G/cm$^2$ \cite{Jochim2003bec}. Throughout the whole evaporation process the magnetic field is set to 1190\,G, above the broad 834-G Feshbach resonance \cite{Inguscio2006ufg,Giorgini2008tou}, where the scattering length between the two $^6$Li spin states is $a=-2900\,a_0$ ($a_0$ is Bohr's radius). Finally, to avoid further evaporative loss, the sample is recompressed by a twofold increase of the trap depth \cite{trapfreqrecomp}.  We obtain a mixture of about $10^{5}$ Li atoms per spin state at a temperature $T^{\rm Li}\approx 100$\,nK together with about $4\times 10^{3}$ K atoms at a temperature $T^{\rm K}\approx 140$\,nK; note that the two species are not fully thermalized at this point, such that $T^{\rm K}>T^{\rm Li}$. In terms of the corresponding Fermi temperatures $T_F^{\rm Li}=650$\,nK and $T_F^{\rm K}=90$\,nK, the temperatures can be expressed as $T^{\rm Li}/T_F^{\rm Li} \approx 0.15$ and $T^{\rm K}/T_F^{\rm K} \approx 1.6$.

This ultracold mixture serves as the starting point for our investigation of collisional loss throughout the crossover of the $^6$Li gas from molecular Bose-Einstein condensation (BEC) to Bardeen-Cooper-Schrieffer (BCS) type behavior \cite{Inguscio2006ufg, Giorgini2008tou,Bloch2008mbp}. This BEC-BCS crossover is controlled by the magnetic field via the broad $^6$Li Feshbach resonance centered at 834\,G. Under our conditions, with a Li Fermi wave number $k^{\rm Li}_F = 1/(4400\,a_0)$, the strongly interacting regime ($|k_F^{\rm Li}a| \geq 1$) is realized between 765 and 990\,G. Note that there are no interspecies Feshbach resonances in the corresponding magnetic field range and the interspecies scattering length remains constantly small at about $+60\,a_0$ \cite{Wille2008eau}. Since there are typically $25$ times less K atoms in the trap than Li atoms, the Li sample remains essentially unperturbed by K.

In a first set of experiments, we ramp the magnetic field to a variable value and study the loss of atoms during a 1-s hold time. For detection, the remaining atoms are recaptured into the two-species MOT and their fluorescence is recorded. The atom numbers determined from fluorescence are calibrated by absorption images.

\begin{figure}
\includegraphics[width=\columnwidth]{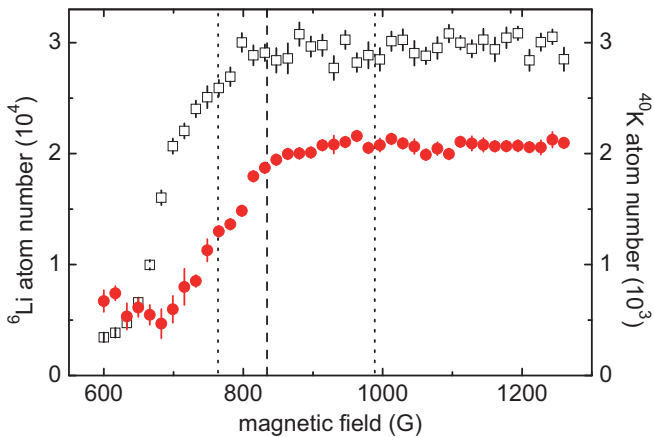}
\caption{\label{fig:Fig1FeshbachScan} Remaining number of $^6$Li atoms per spin state (open squares) and of $^{40}$K atoms (filled circles) after a hold time of 1\,s at a variable magnetic field. The dashed line at 834\,G marks the center of the broad $^6$Li Feshbach resonance and the dotted lines at 765 and 990\,G indicate the strongly interacting regime, where $|k_F^{\rm Li}a| \geq 1$. The error bars indicate the statistical error of ten measurements. }
\end{figure}

The loss of Li and K atoms around the 834-G Li Feshbach resonance is shown in Fig.~\ref{fig:Fig1FeshbachScan}. The Li sample, essentially unperturbed by K, shows a pattern observed previously \cite{Dieckmann2002doa,Bourdel2003mot, JochimPhD}. On the BCS side of the resonance ($\geq834\,$G), Li loss is very small and the detected signal stays nearly constant. On the BEC side, the observed loss results from the vibrational relaxation of dimers during dimer-dimer and atom-dimer collisions \cite{Petrov2004wbm}. The K signal shows a very similar behavior. While loss is very weak on the BCS side of the Li resonance, strong loss is observed on the BEC side with a maximum around 700\,G. At even lower fields, where Li decays very quickly, more K atoms are found to remain. This observation already points to the fact that the main source of K loss results from collisions with Li dimers.

\begin{figure}
\includegraphics[width=\columnwidth]{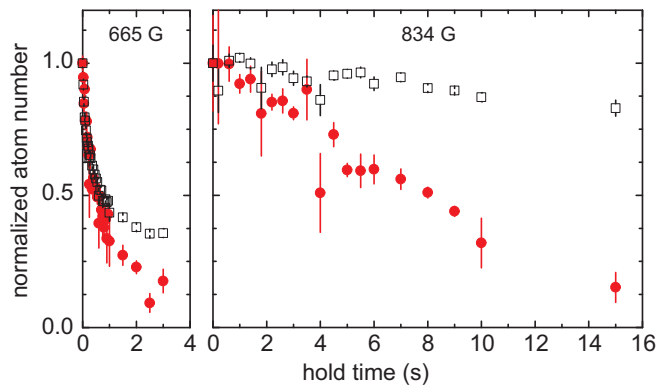}
\caption{\label{fig:Fig2DecayCurves} $^6$Li (open squares) and $^{40}$K (filled circles) atom number normalized to the respective initial atom number over hold time for 665\,G and 834\,G.}
\end{figure}

In a second set of experiments, we study the loss more quantitatively and record the time evolution of the atom numbers for several magnetic field values. Example decay curves are shown in Fig.~\ref{fig:Fig2DecayCurves}. We describe K loss by the rate equation $\dot{N}_{\rm K}/N_{\rm K}=-\Gamma_{\rm bg} - L_2 \left< n_{\rm Li} \right> - L_3 \left< n_{\rm Li}^2 \right>$, where the angle brackets denote averages weighted by the K density distribution. $N_{\rm K}$ is the K atom number, $n_{\rm Li}$ the density distribution of a single Li state and $\Gamma_{\rm bg}$ the background loss rate. The coefficient $L_2$ describes two-body loss resulting from inelastic collisions of K atoms with Li dimers. The coefficient $L_3$ takes into account three-body loss that results from collisions of a K atom with two Li atoms in different spin states. Both loss processes involve three distinguishable atoms, for which there is no Pauli suppression effect as in two-component Fermi systems \cite{Esry2001tlf,Petrov2004wbm}.

For short decay times, the Li density and the K temperature do not change significantly and $-\dot{N}_{\rm K}/N_{\rm K}$ corresponds to an initial decay rate $\Gamma$. We determine $\Gamma$ as the initial slope of an exponential fit to the first 0.5\,s of K decay data (0.3\,s for the 665\,G data point). K and Li density distributions are calculated from measured atom numbers, temperatures, and trap oscillation frequencies~\cite{EndnoteDensityDistributions}. From unitarity to the BCS region, the K cloud is small compared to the Li cloud and essentially probes the center of the Li sample. On resonance, where the peak Li density is $n_{\rm Li,peak}=2.1 \times 10^{12}$\,cm$^{-3}$ for each spin state, we find $n_{\rm Li,peak}/\left< n_{\rm Li} \right>=1.7$ and $n_{\rm Li,peak}^2/\left< n_{\rm Li}^2 \right>=2.4$. For the data point at the lowest magnetic field (665\,G) the Li sample shrinks~\cite{Bartenstein2004cfa} to a size smaller than the K sample and it becomes important to consider the averaged Li density, illustrated by the fact that $n_{\rm Li,peak}/\left< n_{\rm Li} \right>=4$ and $n_{\rm Li,peak}^2/\left< n_{\rm Li}^2 \right>=6.7$.

Based on our experimental data alone we cannot distinguish between two-body or three-body loss processes. We therefore analyze the data by fully attributing the loss to either one of these two processes, after subtracting the measured background loss. This results in coefficients $\mathcal{L}_2=\left(\Gamma-\Gamma_{\rm bg}\right)/\left< n_{\rm Li} \right>$ and $\mathcal{L}_3= \left(\Gamma-\Gamma_{\rm bg}\right)/\left< n_{\rm Li}^2 \right>$, which in general represent upper limits for the real loss coefficients $L_2$ and $L_3$. The background loss rate $\Gamma_{\rm bg}$=0.009\,s$^{-1}$ is determined by analyzing the decay of a pure K sample. Above fields of 900\,G the interspecies loss rate is comparable to this background loss rate.

\begin{figure}
\includegraphics[width=\columnwidth]{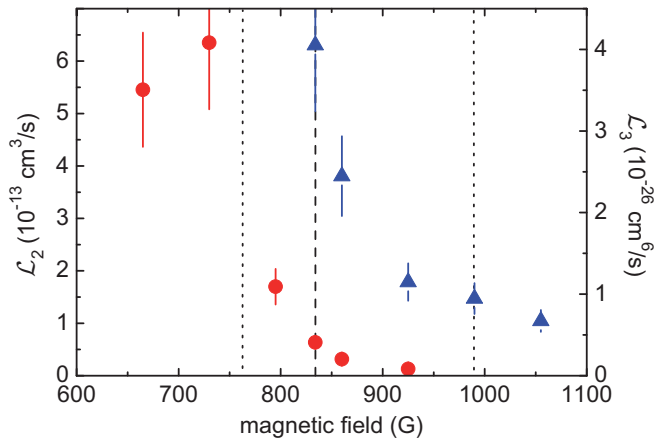}
\caption{\label{fig:Fig3LiKDecayRates} Upper bounds $\mathcal{L}_2$ and $\mathcal{L}_3$ (circles and triangles, respectively) for the two- and three body loss rate coefficients of $^{40}$K immersed in a strongly interacting degenerate $^6$Li sample throughout the $^6$Li BEC-BCS crossover. The dashed line at 834\,G marks the center of the broad $^6$Li Feshbach resonance and the dotted lines at 765 and 990\,G indicate the strongly interacting regime, where $|k_F^{\rm Li}a| \geq 1$. The error bars represent the statistical errors from fitting the loss curves. Additional uncertainties arise from our limited knowledge of the size of the Li sample. On the BCS side this is negligible in comparison with the shown statistical errors, but on the BEC side it may introduce additional errors of up to a few 10\%.}
\end{figure}

The upper bounds $\mathcal{L}_2$ and $\mathcal{L}_3$ for the loss coefficients are shown in Fig.~\ref{fig:Fig3LiKDecayRates}. They peak around 730\,G and decrease by more than an order of magnitude for fields of 925\,G and above. We expect the dominant loss process to change within the $^6$Li BEC-BCS crossover. Far on the BEC side, virtually all Li atoms are bound in Li$_2$ molecules. Potassium is expected to be lost mainly as a result of inelastic two-body collisions with those molecules. At 665\,G the real loss rate coefficient $L_2$ is therefore expected to be very close to $\mathcal{L}_2$, for which we obtain a value of about $6\times 10^{-13}$\,cm$^{-3}/$s. Closer to resonance on the BEC side, atom-dimer loss is substantially reduced in accordance with the predictions of Ref.~\cite{Dincao2008som}.
Far on the BCS side, no Li dimers exist and three-body recombination involving a K$|1\rangle$, a Li$|1\rangle$, and a Li$|2\rangle$ atom can be expected to be the dominant interspecies loss process. At 1190\,G $L_3$ is therefore expected to be very close to $\mathcal{L}_3$, for which we obtain a value of $0.7\times 10^{-26}$\,cm$^{-6}/$s. Near the Feshbach resonance, in the strongly interacting regime, the interpretation of loss is not as straightforward. Three-body loss and an effective two-body loss resulting from collisions with Li pairs may both contribute~\cite{Du2009ico}. At the resonance center (834G) we find $\mathcal{L}_2 = 0.6\times 10^{-13}$\,cm$^{-3}/$s and $\mathcal{L}_3 = 4\times 10^{-26}$\,cm$^{-6}/$s.

Deeper cooling of the Fermi-Fermi mixture leads into the double-degenerate regime \cite{moredetails}. This is achieved by a simple extension of the evaporative cooling process. After completing the forced evaporation ramp (stage I), we hold the sample in the shallow trap for up to 13\,s (stage II) \cite{EndnoteEvaporationToDoubleDegeneracy}; here the Fermi energies are $T_F^{\rm Li} = 420$\,nK and $T_F^{\rm K} = 95$\,nK. In this situation, the Li bath undergoes continuous plain evaporation, which can efficiently remove any heat deposited by the K sample or caused by residual heating processes. Initially the bath of Li atoms contains 3.9$\times 10^4$ per spin state and losses throughout stage II remain below 20\%. The temperature of the Li bath, as determined by fitting Fermi distributions to absorption images of the cloud after trap release, stays constantly at $T^{\rm Li} \approx 40$\,nK.

\begin{figure}
\includegraphics[width=\columnwidth]{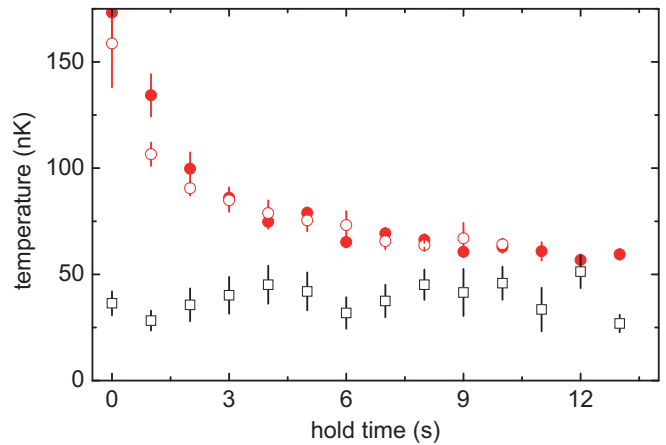}
\caption{\label{fig:Fig4Thermalization} Thermalization of $^{40}$K in a $^6$Li bath. The temperature evolution of the $^{40}$K probe is shown at 1190\,G (filled circles) and at 834\,G (open circles). For comparison we also show the temperature of the $^6$Li bath measured at 1190\,G. The error bars represent the fit errors from analyzing time-of-flight images.}
\end{figure}

Figure \ref{fig:Fig4Thermalization} shows how the K component is cooled down sympathetically into the quantum-degenerate regime. The temperature $T^{\rm K}$ is measured by standard time-of-flight absorption imaging. After the evaporation ramp $T^{\rm K}$ starts at about 170\,nK, as the ramp in stage I is too fast to establish an interspecies thermal equilibrium. In stage II, with the magnetic field kept at 1190\,G (filled circles), the temperature of the $3.4\times10^3$ K atoms slowly approaches a final value of $\sim$$60$\,nK, corresponding to $T^{\rm K}/T^{\rm K}_F = 0.6$. From an exponential fit to $T^{\rm K}$, we derive a $1/e$ time constant for the sympathetic cooling process of $\sim$3\,s. For the Li-K mass ratio, thermalization requires about six collisions \cite{Mudrich2002scw}, giving $\sim$2\,s$^{-1}$ for the elastic collision rate. For comparison, we also study the sympathetic cooling when the magnetic field in stage II is set to 834\,G. In this case, the Li bath forms a strongly interacting superfluid with unitarity-limited interactions \cite{Inguscio2006ufg, Giorgini2008tou}. The only difference observed for the two settings of the magnetic field in stage II is a loss of about 40\% of the K atoms at 834\,G in contrast to an essentially lossless situation at 1190\,G.

Our data in Fig.~\ref{fig:Fig4Thermalization} indicate that the final K temperature stays somewhat above the temperature of the Li bath. This could be readily explained by the presence of a weak heating process of about 5\,nK/s for K counteracting the sympathetic cooling. Another possible explanation could be that we underestimate the Li temperature because of systematic problems with thermometry of deeply degenerate Fermi gases, in particular in our shallow trap with considerable anharmonicities. The final temperatures reached in the sympathetic cooling process need further investigations. In any case, the measurement of the K temperature is free from such systematic errors and thus provides a firm upper bound for the true temperature of the Li bath. Also, a comparison of the scatter and error bars of the data in Fig.~\ref{fig:Fig4Thermalization} for both species highlights that the K thermometry is much less affected by statistical errors, though performed with an order of magnitude less atoms.
These observations point to the great potential of a weakly interacting probe species for precise thermometry in strongly interacting Fermi gases, which in generally is a very difficult task \cite{Inguscio2006ufg, Luo2009tmi, Shin2008pdo}. Note that an impurity thermometer relying on basically the same idea was demonstrated for a Fermi gas of $^{40}$K atoms in Ref.~\cite{RegalPhD}, using a third spin state instead of another species. For stability reasons, however, it was not applied in the strongly interacting regime.

Our ultracold $^{40}$K-$^6$Li combination demonstrates the experimental possibility to immerse another species in a strongly interacting Fermi gas without suffering from collisional loss. With the examples of sympathetic cooling and thermometry we have shown two straightforward applications. There are many more experimental options with the potential to break new ground in research with ultracold fermions. The immersed species can in general serve as a weakly interacting probe for the fermionic superfluid with the great advantages that it can be separately addressed, manipulated, and detected by laser and radio-frequency fields. This may be exploited to study interactions in the many-body regime such as atom-pair collisions or to test the viscosity or the superfluid behavior of the system by a controlled motion of the impurity. Moreover, our double-degenerate system represents an excellent starting point to study a rich variety of phenomena related to few-body quantum states \cite{Petrov2005dmi, Nishida2009cie, Levinsen2009ads} and the rich many-body quantum phases of multicomponent Fermi mixtures \cite{Paananen2006pia, Iskin2006tsf, Iskin2007sai, Petrov2007cpo, Iskin2008tif, Baranov2008spb, Bausmerth2009ccl, Nishida2009ipw, Wang2009qpd, Mora2009gso}.

\begin{acknowledgments}
We acknowledge support by the Austrian Science Fund (FWF) and the European Science Foundation (ESF) within the EuroQUAM/FerMix project and support by the FWF through the SFB FoQuS.
\end{acknowledgments}


\begin{thebibliography}{41}
\bibitem[{tra()}]{trapfreqrecomp}
\bibinfo{note}{The oscillation frequencies for Li (K) are 26\,Hz (10\,Hz)
  axially, and 450\,Hz (260\,Hz) radially, and the trap depth is 3\,$\mu$K
  (6\,$\mu$K).}
\bibitem[{End({\natexlab{a}})}]{EndnoteDensityDistributions}
\bibinfo{note}{The K density distribution is approximated as Gaussian. The Li
  distribution is approximated by a $T$=0 Fermi distribution, rescaled in
  dependence of the magnetic field.}
\bibitem[{End({\natexlab{b}})}]{EndnoteEvaporationToDoubleDegeneracy}
\bibinfo{note}{In stage II, the trap oscillation frequencies for Li (K) are
  26\,Hz (10\,Hz) axially, and 330\,Hz (190\,Hz) radially. The trap depth is
  1.5\,$\mu$K (3\,$\mu$K).}
\end{thebibliography}
\end{document}